# Reference-Point Dependence in Multiplicative Kantian Optimization

Igor Sloev[1 2], Gerasimos Lianos[3]

This note shows that multiplicative Kantian optimization is reference-point dependent. In the standard formulation, the Multiplicative Kantian Equilibrium (MKE) test is applied to absolute contributions, and Nash optimizers may free-ride on Kantians. We compare this with a formulation in which Kantians apply the same multiplicative test to contributions above the individually rational Nash level. With the reference point at the Nash contribution, the Nasher's action is zero in the auxiliary scale. Since zero is a fixed point of every multiplicative transformation, the Kantian test makes the Nash outcome stable in mixed encounters. In a public-good game with a dominant Nash strategy, this neutralizes the free-riding advantage of Nash optimizers, while two Kantians attain the efficient symmetric MKE under the efficient-equilibrium selection convention. The induced type-payoff matrix makes the Kantian type weakly dominant and evolutionarily stable. Thus, the stability of Kantian behavior depends not only on the multiplicative Kantian criterion, but also on the reference point used to formulate it.

[1]School of Computational Social Sciences, European University at St. Petersburg, Russia, ORCID: 0000-0003-0286-1034, sia@eu.spb.ru.

[2]Corresponding author: 6/1A Gagarinskaya st., St. Petersburg 191187 Russia, sia@eu.spb.ru.
[3]Higher Colleges of Technology, Abu Dhabi College, Baniyas, Abu Dhabi, UAE, glianos@hct.ac.ae

## 1. Introduction

John Roemer (2010, 2015, 2019) introduced the concept of Kantian optimization, in which an agent evaluates a strategy profile by asking: "What if everyone scaled their action by the same factor?" rather than the Nash question: "What if I alone change my action?" This idea echoes Kant's categorical imperative — act only according to a maxim that you could will to become a universal law — translated here into a concrete behavioral rule: multiply all actions by a common number. Under appropriate conditions, the MKE can select Pareto-efficient outcomes, in contrast to the inefficient Nash equilibrium. The MKE concept has been applied to a range of problems, including public-good provision, common-pool resource management, and team production.

One challenge for Kantian optimization is its vulnerability to free-riding. If some players remain Nash optimizers (Nashers) while others adopt the Kantian criterion, the Nashers may obtain higher payoffs than the Kantians. This observation has motivated a debate on whether Kantian optimization can be viewed as rational or whether it necessarily relies on non-material motivations (Roemer 2020; Sher 2020; Braham and Van Hees 2020).

This note investigates whether a Kantian can avoid exploitation by Nashers while retaining the Kantian criterion in all interactions, without switching to Nash optimization against non-cooperative opponents. We show that the answer depends on how the Kantian maxim is formulated. Specifically, we compare two simple versions of the same public-good game. In the first, Kantians apply the multiplicative test to absolute contributions. In the second, they apply it to the part of the contribution that exceeds the individually rational Nash level. The two formulations produce sharply different outcomes: in the first, Nashers free-ride and earn more than Kantians; in the second, free-riding is neutralized, and the Kantian type becomes evolutionarily stable.

The underlying mechanism is simple. The multiplicative Kantian test compares a profile with all profiles obtained by multiplying every action by a common factor $a$. Hence the zero of the scale in which actions are measured is a fixed point: $a \cdot 0 = 0$ for any $a$. The set of profiles that pass the test therefore depends on which action is taken as the reference point of the scale. We call this property reference-point dependence. When the reference point is at the Nash contribution, the Nash profile $(x^N, x^N)$ becomes immune to Kantian deviations in mixed encounters, eliminating the free-rider advantage. When the reference point is at zero contribution, free-riding persists.

Our contribution is twofold. First, we demonstrate that the stability of Kantian optimization in a mixed population is not an intrinsic property of the Kantian criterion alone, but depends on the reference point for the universalization test. Second, we show that a natural reference point — the non-cooperative Nash action — yields a type-payoff matrix in which the Kantian type weakly dominates the Nash type and is evolutionarily stable. Crucially, this result does not require Kantians to change their optimization protocol when facing a Nasher, unlike the framework of Curry and Roemer (2012) and Grafton et al. (2017).

The remainder of the note is organized as follows. Section 2 introduces the public-good game and defines the two scenarios for the Kantian criterion. Section 3 solves for equilibrium outcomes in pairwise encounters and derives the payoff matrices. Section 4 applies the matrices to type choice and evolutionary stability. Section 5 concludes.

## 2. Model

There are two players, each of type Kantian ($K$) or Nasher ($N$). The material payoff of player $i$ is

$$u_i(x_i, x_j) = b(x_i + x_j) - c(x_i), \quad x_i \in [0, \bar{x}],$$

with $b > 0, c(0) = 0, c' > 0, c'' > 0$ on $[0, \bar{x}]$, and $c'(0) < b < 2b < c'(\bar{x})$.

A Nasher maximizes $u_i(x_i, x_j)$ over $x_i \in [0, \bar{x}]$ given $x_j$. The unique best response, defined by $c'(x^N) = b$, is strictly dominant. The symmetric Nash payoff is $U^N = u(x^N, x^N)$.

The efficient symmetric outcome $x^P$ maximizes $u_1(x, x) + u_2(x, x)$ along the diagonal, giving $c'(x^P) = 2b$, with $x^P > x^N$ and $U^P = u(x^P, x^P) > U^N$.

In a mixed *K*–*N* encounter, an equilibrium is a profile $(x_K, x_N)$ such that $x_N$ is a Nash best response and the profile satisfies the Kantian's MKE condition in the relevant scenario.

## 3. Two Scenarios for the Kantian Maxim

Since $c'' > 0$, the one-dimensional payoff functions generated by the scaling tests below are strictly concave whenever the scaled action is nonzero; hence the first-order conditions are sufficient. We compare two formulations of the multiplicative Kantian test. They differ in the reference point from which actions are scaled. To save notation, whenever we write $\forall a \geq 0$, we mean all nonnegative $a$ for which the scaled profile remains feasible in the relevant scenario.

### 3.1. Scenario A: reference point at zero

The Kantian applies the MKE test directly to absolute contributions. Admissible actions: $x_i \in [0, \bar{x}]$. A profile $(x_1, x_2)$ is MKE if

$$u_i(x_1, x_2) \geq u_i(ax_1, ax_2) \quad \forall a \geq 0.$$

A *K–K* pair has two symmetric MKEs: 0 and $x^P$. Selecting the efficient one gives both payoffs $U^P$. In a *K–N* encounter the Nasher plays $x^N$. The Kantian's MKE condition requires that $a = 1$ maximizes

$$f(a) = u_1(ax_1, ax^N) = ba(x_1 + x^N) - c(ax_1).$$

The first-order condition $f'(1) = 0$ gives

$$x_1(c'(x_1) - b) = bx^N. \tag{2}$$

Since the right-hand side is positive, $c'(x_1) > b = c'(x^N)$. With $c'' > 0$ this implies $x_1 > x^N$. Evaluating the left-hand side at $x_1 = x^P$ (where $c'(x^P) = 2b$) gives $x^P b > bx^N$, so the solution satisfies $x_1 < x^P$. Hence $x^N < x_1 < x^P$.

The resulting payoffs are

$$u_K^A = b(x_1 + x^N) - c(x_1), \quad u_N^A = b(x_1 + x^N) - c(x^N).$$

Because $x_1 > x^N$ and c is strictly increasing,

$$u_N^A - u_K^A = c(x_1) - c(x^N) > 0.$$

Moreover, using $c(x_1) - c(x^N) > c'(x^N)(x_1 - x^N) = b(x_1 - x^N)$,

$$U^N - u_K^A = c(x_1) - c(x^N) - b(x_1 - x^N) > 0,$$

while $u_N^A - U^N = b(x_1 - x^N) > 0$. Thus $u_K^A < U^N < u_N^A$.

Free-riding prevails: the Nasher earns strictly more than the Kantian, and strictly more than the symmetric Nash payoff.

### 3.2. Scenario B: reference point at $x^N$

Here the Kantian views cooperation as contributing above the egoistic level. Admissible actions are $x_i \in [x^N, \bar{x}]$. The material payoff remains defined on the original action space; the restriction concerns the cooperative domain over which the Kantian test is formulated.

Define deviations $z_i = x_i - x^N \geq 0$ and auxiliary payoff $v_i(z_1, z_2) = u_i(z_1 + x^N, z_2 + x^N)$. A profile $(z_1, z_2)$ is MKE if

$$v_i(z_1, z_2) \geq v_i(az_1, az_2) \quad \forall a \geq 0.$$

*K–K interaction.* With symmetry $z_1 = z_2 = z \geq 0$, the MKE condition requires that $a = 1$ maximizes

$$F(a) = v_i(az, az) = 2b(az + x^N) - c(az + x^N).$$

The first-order condition yields two solutions: $z = 0$ and $z = x^P - x^N > 0$. Both are symmetric MKE. We select the efficient MKE, giving contributions $x^P$ and payoffs $U^P$.

*K–N interaction.* The Nasher plays $x^N$, so $z_2 = 0$. The MKE condition becomes

$$u_1(z_1 + x^N, x^N) \geq u_1(az_1 + x^N, x^N) \quad \forall a \geq 0.$$

Since $a \cdot 0 = 0$, the Nasher's action stays fixed. The test collapses to checking whether $z_1$ is a best response to $x^N$. The unique best response is $x^N$ itself, so $z_1 = 0$. Any $z_1 > 0$ is not MKE: choosing $a = 0$ gives a higher payoff. Hence the unique *K–N* outcome is $z_1 = z_2 = 0$, i.e., $x_1 = x_2 = x^N$, with both obtaining $U^N$. No free-riding occurs.

### 3.3. Comparison

The two scenarios produce different payoff matrices. Under Scenario A, the *K*–*N* cell has $u_K^A < U^N < u_N^A$. Under Scenario B, the matrix is:

*K*–*K* interaction: both receive $U^P$,

*K*–*N* interaction: both receive $U^N$,

*N*–*N* interaction: both receive $U^N$.

Here $U^P > U^N$. The only selection convention is the efficient-MKE choice in the *K*–*K* cell; the *K*–*N* outcome is unique.

The difference is driven by a single property: $a \cdot 0 = 0$ for any $a$. In Scenario B the reference point is $x^N$, so the Nasher's action $z_2 = 0$ is invariant under scaling, and the Kantian test collapses to a best-response calculation at $x^N$. In Scenario A the reference point is 0, so scaling moves the Nasher's action away from $x^N$, and a different outcome emerges. We call this property *reference-point dependence*: the set of MKE outcomes depends on which action serves as the reference point for the multiplicative test.

## 4. Implications: Type Choice and Evolutionary Stability

We now take the Scenario B payoff matrix derived above as the induced type game. Recall that the only selection convention is the efficient-MKE choice in the *K*–*K* cell; the *K*–*N* outcome is unique.

$$\pi(K,K) = U^P, \qquad \pi(K,N) = \pi(N,K) = \pi(N,N) = U^N, \text{ where } U^P > U^N.$$

Let $k \in [0,1]$ be the fraction of Kantians. With observable types and random matching,

$$U_K(k) = kU^P + (1-k)U^N, \qquad U_N(k) = U^N.$$

Thus $U_K(k) \geq U_N(k)$, with strict inequality for $k > 0$. Choosing Kantian weakly dominates choosing Nasher, and the all-Kantian profile is an equilibrium delivering $U^P$ in every match. The all-Nasher profile is also an equilibrium, but it is not robust to any positive mass of Kantians.

Under the standard ESS definition, Kantian is a strict ESS because

$$\pi(K,K) = U^P > U^N = \pi(N,K).$$

Nasher is not ESS because $\pi(K,N) = \pi(N,N) = U^N$, while $\pi(K,K) > \pi(N,K)$. Hence the Kantian type is evolutionarily stable, whereas the Nasher type is not.

## 5. Concluding Remarks

This note has shown that the set of outcomes under multiplicative Kantian optimization depends on which action serves as the reference point for the Kantian counterfactual. We compared two natural formulations of the Kantian maxim in a simple public-good game with a dominant Nash strategy. When the reference point is at zero contribution, the familiar free-riding outcome arises: a Nasher obtains more than a Kantian in a mixed encounter. When the reference point is at the individually rational Nash contribution, the free-riding advantage vanishes: the Nash profile becomes Kantian-stable in mixed encounters, while two Kantians continue to attain the efficient outcome. The resulting payoff matrix makes the Kantian type weakly dominant and evolutionarily stable.

The underlying mechanism — reference-point dependence — follows directly from the fact that the multiplicative test fixes the zero of the scale. This property is not specific to the game studied here; it is a general feature of multiplicative Kantian equilibria.

We leave the endogenous choice of reference points for future work.

## References

Braham, M., Van Hees, M. (2020). Kantian optimization, rationality, and the categorical imperative. Economics and Philosophy, 36(3), 436–453.

Curry, P.A., Roemer, J.E. (2012). Kantian optimization and the stability of cooperation. Journal of Public Economic Theory, 14(4), 609–631.

Grafton, R.Q., Kompas, T., Van Long, N. (2017). A brave new world? Kantian–Nashian interaction and the dynamics of global climate change mitigation. European Economic Review, 99, 31–42.

Roemer, J.E. (2010). Kantian equilibrium. Scandinavian Journal of Economics, 112(1), 1–24.

Roemer, J.E. (2015). Kantian optimization: a microfoundation for cooperation. Journal of Public Economics, 127, 45–57.

Roemer, J.E. (2019). How We Cooperate: A Theory of Kantian Optimization. Yale University Press.

Roemer, J.E. (2020). Kantian optimization, rationality, and the categorical imperative: response to Sher. Economics and Philosophy, 36(3), 455–464.

Sher, I. (2020). Kantian optimization, rationality, and the categorical imperative: a comment on Roemer. Economics and Philosophy, 36(3), 425–435.